\documentclass[
	reprint,
    superscriptaddress,
    amsmath,
    amsfonts,
    amssymb,
    aps,
    prl]{revtex4-1}

\usepackage{physics}
\usepackage{graphicx}
\usepackage[hyperindex,breaklinks]{hyperref}
\hypersetup{
	pdfstartview={FitH},
	pdfnewwindow=true,
	colorlinks=true,
	linkcolor=blue,
	citecolor=blue,
	filecolor=blue,
	urlcolor=blue}
\usepackage[capitalize, nameinlink]{cleveref} 

\newcommand{\amp}{\mathfrak{amp}}
\newcommand{\comp}{\mathfrak{comp}}
\newcommand{\rot}{\mathfrak{rot}}
\newcommand{\unif}{\mathfrak{unif}}

\newcommand{\MQ}{%
	\affiliation{%
        Department of Physics and Astronomy,
        Macquarie University,
        Sydney NSW, Australia.}}
\newcommand{\QuArC}{%
    \affiliation{%
        Quantum Architectures and Computation Group,
        Microsoft Research,
        Redmond WA, United States of America.}}
        
\begin{document}

\title{Black-box quantum state preparation without arithmetic}

\author{Yuval R.~Sanders} \MQ
\author{Guang Hao Low} \QuArC
\author{Artur Scherer} \MQ
\author{Dominic W.~Berry} \MQ

\date{\today}

\begin{abstract}
Black-box quantum state preparation is an important subroutine in many quantum 
algorithms. The standard approach requires the quantum computer to do 
arithmetic, which is a key contributor to the complexity. Here we present a new 
algorithm that avoids arithmetic. We thereby reduce the number of gates by a 
factor of $286$--$374$ over the best prior work for realistic precision; the 
improvement factor increases with the precision. As quantum state preparation is 
a crucial subroutine in many approaches to simulating physics on a quantum 
computer, our new method brings useful quantum simulation closer to reality.
\end{abstract}

\maketitle


Black-box quantum state preparation, first developed by Grover~\cite{Gro00} as 
an extension of his more famous search algorithm~\cite{Gro97}, is a widely used 
quantum computational primitive. State preparation is invoked, for example, in 
the discrete-time quantum walk approach to Hamiltonian 
simulation~\cite{Chi09,BC12,LC17}, and in the linear combination of unitaries 
(LCU) technique~\cite{Kot14,BCK15,BCC+15,CKS17,LC16}. State preparation 
procedures are also treated as input to quantum algorithms for solving systems
of linear equations~\cite{HHL09,CKS17}, data fitting~\cite{WBL12}, and 
computing scattering cross sections~\cite{CJS13,SVM+17}.

Grover gave a black-box state preparation algorithm that requires the quantum 
computer to calculate arcsines, which is a major contributor to the complexity. 
Here we eliminate the need for arithmetic and thereby greatly improve the cost. 
For instance, Grover's approach would require over $11000$ Toffoli 
gates~\cite{HRS18} per round of amplitude amplification to prepare a quantum 
state to $30$ bits of precision. Our approach requires only $30$ Toffoli gates. 
This complexity reduction brings useful Hamiltonian simulation much closer to 
reality.

The scenario for black-box state preparation is as follows. We are given access 
to a quantum oracle $\amp$ (the `black box') that returns target coefficients as 
follows: if $\vec{\alpha} := \left( \alpha_0, \alpha_1, \ldots, \alpha_{d-1} 
\right)$ is a real vector with $0 \leq \alpha_\ell < 1$ for each $\ell = 0, 
\ldots, d-1$, then
\begin{equation} \label{eq:oracle}
	\amp \ket{\ell} \ket{z} :=
	\ket{\ell}  \ket*{z \oplus \alpha_\ell^{(n)}},
\end{equation}
where $z$ is an $n$-bit integer encoded into an $n$-qubit register, $\oplus$ 
represents a bitwise XOR, and $\alpha_\ell^{(n)} = \left\lfloor 2^n \alpha_\ell 
\right\rfloor$. The task is to prepare an approximation to the `target' state 
\begin{equation}
  \ket{\text{target}} :=
  \frac{1}{\|\vec{\alpha}\|_2}
  \sum_{\ell=0}^{d-1} \alpha_\ell \ket{\ell}.
\end{equation}

A key contributor to the cost of Grover state preparation is the use of a 
subroutine defined as 
\begin{equation}
	\rot \ket{\xi} \ket{0} :=
    	\ket{\xi} \left( \sin\theta \ket{0} + \cos\theta \ket{1} \right),
\end{equation}
where $\xi$ is one of the values $\alpha_\ell^{(n)}$, the second register is a 
qubit and $\theta$ is a high-precision approximation to $\arcsin \left( \xi/2^n 
\right)$. To implement this procedure, the quantum computer would calculate 
$\theta$, store the value in an ancillary register, and use that register as the 
control for a sequence of rotation operations on the qubit. Grover then 
prescribes roughly $\sqrt{d} / \norm{\vec\alpha}_2$ rounds of amplitude 
amplification in which both $\amp$ and $\rot$ are invoked. The number of rounds 
of amplitude amplification, and hence the number of uses of $\amp$, is provably 
optimal~\cite{BHMT02}.

In this Letter, we eliminate $\rot$ and replace it with a new procedure defined 
as
\begin{equation}
	\comp \ket{a}\ket{b}\ket{0} := 
    	\left\{
        	\begin{matrix}
        		\ket{a} \ket{b} \ket{0} & \text{if } a < b, \\
        		\ket{a} \ket{b} \ket{1} & \text{if } a \geq b,
			\end{matrix}
        \right.
\end{equation}
where $a$ and $b$ are $n$-bit integers. A similar approach was used 
in~\cite{BGB+18} (Sec.~III.D) for state preparation in an LCU-based quantum 
simulation algorithm, though the task was different. In that paper, basis states 
are allowed to be entangled with ancillae. Moreover, the amplitudes were 
recorded in a classical database rather than provided as the output of a quantum 
oracle.


To explain our method, we first review Grover's approach in more detail. In 
Grover state preparation, the following three operations are executed in 
sequence.
\begin{enumerate}
    \item \emph{Prepare the output register.}
    The output register, which we call \texttt{out}, is a $d$-dimensional
    quantum register that, like all registers, starts in the zero state.
    We prepare \texttt{out} in a uniform superposition of all computational
    basis states. 
    
    \item \emph{Write target amplitudes.}
    Use $\amp$ to write the target amplitude to a new $n$-qubit register
    called \texttt{data} controlled by the value in \texttt{out}.
    
    \item \emph{Amplitude transduction.}
    Apply the procedure $\rot$, which performs a rotation
    on a single qubit called \texttt{flag} controlled by the arcsine of
    the value in \texttt{data}.
\end{enumerate}
Grover's black-box state preparation algorithm prescribes $\order{ \sqrt{d} / 
\norm{\vec\alpha}_2 }$ rounds of amplitude amplification on the above steps, 
where the register \texttt{flag} indicates the correct subspace. Following 
amplitude amplification, reset the \texttt{data} register by applying $\amp$ 
once more; \texttt{data} can then be discarded. Finally, measure \texttt{flag}. 
If the result is one, the algorithm has failed. Otherwise, the state of the 
register \texttt{out} is approximately $\ket{\text{target}}$ as required.

The key step of Grover's black-box state preparation algorithm is amplitude 
transduction, in which the value $\alpha_\ell^{(n)}$ recorded in the 
\texttt{data} register is transduced into an amplitude for 
$\ket{\ell}_\texttt{out}$. In Grover's approach, the procedure $\rot$ is used to 
perform amplitude transduction. The state of the quantum computer after applying 
$\rot$ but before amplitude amplification is
\begin{multline}
    \ket{\psi_\text{rot}} :=
    \frac{1}{\sqrt{d}} \sum_{\ell=0}^{d-1}
    \ket{\ell}_\texttt{out}
    \ket*{\alpha_\ell^{(n)}}_\texttt{data} \\
    \otimes
    \left( \sin \theta_\ell \ket{0}_\texttt{flag} +
           \cos \theta_\ell \ket{1}_\texttt{flag} \right),
\end{multline}
where $\theta_\ell \approx \arcsin(\alpha_\ell^{(n)}/2^n)$. The amplitude $\sin 
\left( \theta_\ell \right) \approx \alpha_\ell$ is thus applied to the 
appropriate part of the superposition in the subspace marked by 
$\ket{0}_\texttt{flag}$.

In our approach, we instead perform amplitude transduction by testing an 
inequality. We compare the value of the oracle output with a superposition of 
all possible outputs prepared in a new $n$-qubit register called \texttt{ref} 
with the result written to \texttt{flag}. We call this operation $\comp$. The 
state after applying $\comp$ is
\begin{multline} \label{eq:comp}
    \ket{\psi_\text{comp}} := 
    \frac{1}{\sqrt{2^n d}}
    \sum_{\ell=0}^{d-1}
    \ket{\ell}_\texttt{out}
    \ket*{\alpha_\ell^{(n)}}_\texttt{data} \\
    \otimes \left(
        \sum_{x=0}^{\alpha_\ell^{(n)} - 1}
            \ket{x}_\texttt{ref} \ket{0}_\texttt{flag} + 
        \sum_{x=\alpha_\ell^{(n)}}^{2^n - 1}
            \ket{x}_\texttt{ref} \ket{1}_\texttt{flag}
    \right).
\end{multline}
Our next step is to unprepare the uniform superposition on \texttt{ref} with $n$ 
Hadamard gates. That yields the state
\begin{multline}
	\frac{1}{\sqrt{d}} \sum_{\ell=0}^{d-1}
    \frac{\alpha_\ell^{(n)}}{2^n} \ket{\ell}_\texttt{out}
    \ket*{\alpha_\ell^{(n)}}_\texttt{data}
    \ket{0}^{\otimes n}_\texttt{ref} \ket{0}_\texttt{flag} \\
    + \ket{\omega}_{%
        \texttt{out}
        \otimes
        \texttt{data}
        \otimes
        \texttt{ref}
        \otimes
        \texttt{flag}},
\end{multline}
where $\ket{\omega}$ is an unnormalized state containing the parts of the 
superposition with non-zero values encoded on $\texttt{ref} \otimes 
\texttt{flag}$. The remainder of the algorithm works as before, except that 
success is indicated by zero on both \texttt{flag} and \texttt{ref} instead of 
only \texttt{flag} as before.


We can extend our approach to the case in which we want to prepare a state that 
is approximately proportional to $\sum_\ell \sqrt{\alpha_\ell} \ket{\ell}$, 
rather than $\sum_\ell \alpha_\ell \ket{\ell}$. This variant of state 
preparation is used, for example, in the discrete-time quantum walk approach to 
Hamiltonian simulation~\cite{Chi09, BC12}. Notice that the number of terms 
marked by $\ket{0}_\texttt{flag}$ in \cref{eq:comp} is $\alpha_\ell^{(n)}$, 
meaning that the amplitude on the subspace marked by $\ket{\ell}_\texttt{out} 
\ket{0}_\texttt{flag}$ is proportional to $\sqrt{\alpha_\ell^{(n)}}$. We want to 
preserve these square-root amplitudes whilst resetting \texttt{ref}. To do this, 
we simply skip the step where we unprepare a uniform superposition on 
\texttt{ref}. Following amplitude amplification to boost the amplitude of 
$\ket{0}_\texttt{flag}$, we reset the \texttt{ref} register using a new 
procedure called $\unif^{-1}$. This new procedure is the inverse of $\unif$, 
which is defined so that
\begin{equation} \label{eq:unif}
	\unif \ket{\Lambda}_\texttt{data} \ket{0}_\texttt{ref} :=
    	\frac{1}{\sqrt{\Lambda}} \ket{\Lambda}_\texttt{data}
        \sum_{x=0}^{\Lambda-1} \ket{x}_\texttt{ref}.
\end{equation}
Note that we do not reset \texttt{ref} during amplitude amplification, meaning 
that we apply $\unif^{-1}$ only once at the end. This introduces an additive, 
rather than multiplicative, complexity overhead to the algorithm. This also 
means that any inaccuracy in $\unif^{-1}$ is introduced only once. It therefore 
suffices to explain how to execute $\unif$ to within an error that is 
proportional to the overall error tolerance of the algorithm.

The operation $\unif$ can be approximated to within $\varepsilon$ using 
fixed-point amplitude amplification (Theorem~27 in~\cite{GSLW18}) on an 
intermediate procedure $\unif^\prime$ defined so that
\begin{multline}
	\unif^\prime \ket{\Lambda}_\texttt{data}
    \ket{0}_\texttt{ref} \ket{0}_\texttt{anc} := \\
		\frac{1}{\sqrt{\Lambda^\prime}} \ket{\Lambda}_\texttt{data} \left(
        	\sum_{x=0}^{\Lambda-1}
            \ket{x}_\texttt{ref} \ket{0}_\texttt{anc} +
        	\sum_{x=\Lambda}^{\Lambda^\prime - 1}
            \ket{x}_\texttt{ref} \ket{1}_\texttt{anc}
        \right),
\end{multline}
where $\Lambda^\prime := 2^{\left\lceil \log_2 \Lambda \right\rceil}$ and 
\texttt{anc} is an ancillary qubit used to mark the desired state. Note that the 
desired output is marked by $\ket{0}_\texttt{anc}$ with amplitude 
$\sqrt{\Lambda/\Lambda^\prime} > 1/\sqrt{2}$. Also note that $\unif^\prime$ can 
be executed using $n$ controlled-Hadamard gates followed by an application of 
$\comp$. We may then use fixed point amplitude amplification (set $a = 1$ and 
$\delta = 1 / \sqrt{2}$ in Theorem~27 of~\cite{GSLW18}) to boost the amplitude 
of $\ket{0}_\texttt{anc}$ to within $1 - \varepsilon$. The cost of this approach 
is $\order{n \log (1/\varepsilon)}$.


We can also extend our approach to prepare states with complex amplitudes. First 
we briefly establish some terminology. We have explained above how to prepare 
states whose amplitudes are equal to either the output of an oracle or the 
square root of that output. We call the first state preparation problem the 
`linear coefficients' problem and the second the `root coefficients' problem. 
The original form of both the linear and root coefficients problems assumes that 
the target amplitudes are positive and real; we can generalize to complex 
coefficients that are presented in either polar form or Cartesian form. That is 
to say, the amplitudes are presented to us as the output of two oracles, rather 
than one. For polar form, these oracles return either the magnitude or the 
argument of the target amplitudes; for Cartesian form, the oracles return either 
the real or imaginary component.

To solve the polar form of either the linear or root coefficients problem, first 
prepare a state approximately proportional to $\sum_\ell \abs{\alpha_\ell} 
\ket{\ell}$ or $\sum_\ell \sqrt{\abs{\alpha_\ell}} \ket{\ell}$, respectively, as 
before using the magnitude oracle in place of $\amp$. Then use the argument 
oracle to write $\arg \left( \alpha_\ell \right)$ to \texttt{data}. We then 
simply perform a sequence of controlled phase operations (controlled by 
\texttt{data}) to transduce the phases to \texttt{out}; the phase in the 
controlled phase operations depends on whether we seek to solve the linear or 
root coefficients problem.

Solving the Cartesian form of each problem is more complicated. We describe the 
linear and root coefficients problems separately. For the linear coefficients 
problem, we extend the \texttt{flag} register to two qubits.  We initialize the 
new \texttt{flag} qubit to $\left(\ket{0} + i \ket{1}\right)/\sqrt{2}$. We then 
replace the oracle queries in our algorithm with applications of both the real 
and imaginary oracle conditioned on the value of the new \texttt{flag} qubit 
($\ket{0}$ for real, $\ket{1}$ for imaginary). As the real and imaginary parts 
may also be negative, each oracle is assumed to return a signed integer instead 
of unsigned as before; we transduce the sign to an amplitude by applying the $Z$ 
gate to the sign bit. After applying $\comp$ and then resetting \texttt{data} 
with another conditional application of the real or imaginary oracle, we apply a 
Hadamard gate to the new \texttt{flag} qubit. The desired state is now flagged 
by $\ket{0}^{\otimes 2}_\texttt{flag}$ and we can perform amplitude 
amplification as usual.

For the Cartesian form of the root coefficients problem, we cannot entirely 
avoid arithmetic, although we can keep it to a minimum. We follow the same 
procedure as for the Cartesian form of the linear coefficients problem, but we 
modify the application of $\comp$. To explain this modification, first note the 
following. If we set $\alpha_\ell = a_\ell + i b_\ell$, observe that $r := 
\operatorname{Re} \left( \sqrt{\alpha_\ell} \right)$ satisfies $4 r^2 \left( r^2 
- a_\ell \right) = b_\ell^2$ and that $c := \operatorname{Im} \left( 
\sqrt{\alpha_\ell} \right)$ satisfies $4 c^2 \left( c^2 + a_\ell \right) = 
b_\ell^2$. For the principal root of $\alpha_\ell$ (the one we seek to prepare), 
the real part is non-negative and the sign of the imaginary part matches that of 
$b_\ell$. We then modify the application of $\comp$ as follows: instead of using 
$\comp$ to compare the value of the oracle output and a superposition of 
possible amplitudes $x$, we test whether $4 x^2 \left( x^2 \pm a_\ell \right) < 
b_\ell^2$. The choice of $+$ or $-$ is made controlled on the state of the new 
\texttt{flag} qubit. This new step requires a handful of arithmetic operations, 
though fortunately we need not calculate a square root directly.


Having explained our new approach to black-box quantum state preparation, we now 
analyze its complexity. We focus on analyzing the complexity of our approach for 
preparing positive amplitudes; the analysis of complex amplitude preparation is 
similar. To assess the cost of our algorithm, we count only non-Clifford gates, 
which are considered far more costly than Clifford gates due to the overhead 
incurred in fault-tolerant execution~\cite{FMMC12}. During each round of 
amplitude amplification, the only non-Clifford operations are $\comp$ and 
$\amp$. There are also Hadamard operations performed on \texttt{out} and 
\texttt{ref}, but those are Clifford operations. The operation $\amp$ is 
regarded as a black box, which means we count the number of its uses but do not 
attempt to count gates required to implement it. Each round of amplitude 
amplification uses $\amp$ and its inverse once, then $\amp$ is used once after 
amplitude amplification to reset \texttt{data} (the same as for Grover's 
approach).

Our improvement in complexity comes about because $\comp$ can be performed 
efficiently. As shown in~\cite{Gid18}, $\comp$ can be executed using $n$ Toffoli 
gates (non-Clifford) and $n$ additional qubits. Together with $n$ additional 
qubits for each of the \texttt{data} and \texttt{ref} registers and one 
additional \texttt{flag} qubit, we have a total cost of $n$ non-Clifford gates 
per round of amplitude amplification and an overall space cost of $3n+1$ qubits. 
One could instead implement $\comp$ using the techniques of~\cite{CDKM04} with 
only $2n+2$ qubits, but then $2n-1$ non-Clifford gates would be required.

The amplitude of the desired state before amplitude amplification in the linear 
coefficients problem is approximately $\norm{\vec\alpha}_2/\sqrt{d}$. Therefore 
the number of rounds of amplitude amplification is $\approx \frac \pi 4 
\sqrt{d}/\norm{\vec\alpha}_2$ (which is the same as for Grover's approach). 
Similarly, we need $\approx \frac \pi 4 \sqrt{d/\norm{\vec\alpha}_1}$ rounds for 
the root coefficients problem. Each round of amplitude amplification uses 
$\comp$ and the oracle $\amp$ twice.

After amplitude amplification is complete, we must reset the \texttt{data} 
register, which takes another application of $\amp$. For the linear coefficients 
problem no further non-Clifford gates are required. For the root coefficients 
problem, $\order{n \log (1/\varepsilon)}$ non-Clifford gates are used to apply 
$\unif^{-1}$ with $\varepsilon$ precision according to Theorem~27 
in~\cite{GSLW18}. Thus the complexity of our black-box state preparation 
approach is, in terms of Toffoli gates, $\frac \pi 2 n \sqrt{d} / 
\norm{\vec\alpha}_2 + \order{1}$ for the linear coefficients problem and $\frac 
\pi 2 n \sqrt{d/\norm{\vec\alpha}_1} + \order{n \log(1/\varepsilon)}$ for the 
root coefficients problem.

Having established the complexity of our algorithm, we now compare with Grover. 
As we use the same number of rounds of amplitude amplification, our improvement 
arises from the elimination of the use of $\rot$. To execute $\rot$, Grover uses 
a sequence of conditional rotations by an angle calculated as the arcsine of the 
value stored in the \texttt{data} register. The number of conditional rotations 
depends on the precision of the arcsine calculation; we assume that this is $n$ 
bits. The cost of $\rot$ is therefore equal to $n$ controlled qubit rotations 
(each of which is non-Clifford) together with the cost of calculating the 
arcsine of the value stored in \texttt{data}.

The most thorough cost analysis for calculating an arcsine on a quantum computer 
is given in Appendix~D.2 of~\cite{HRS18}, which reports a complexity of 
$\Omega(n^2)$. That is to say, the complexity is $k n^2$ for a value $k$ that 
increases with $n$. Table~II of~\cite{HRS18} presents explicit complexity values 
for specific target accuracy values of $10^{-5} \approx 2^{-17}$, $10^{-7} 
\approx 2^{-23}$, and $10^{-9} \approx 2^{-30}$. We can roughly compare these 
numbers to the cost of our algorithm for $n = 17$, $23$, and $30$, respectively. 
We present this comparison in Table~\ref{tab:compare_to_Grover}. Note that we 
are not counting the cost of non-Clifford gates for conditional rotations, so 
are underestimating the cost of Grover's approach.

These gate counts are for operations that are performed twice per round of 
amplitude amplification, but the number of rounds is unchanged. The improvement 
factor is exactly the same for the entire algorithm provided we consider the 
linear coefficients problem. For the root coefficients problem there are other 
operations needed, but it can be expected that there will be fewer of those so 
the improvement factor will be similar. Our approach provides an improvement by 
a factor of $286$ to $375$ for the precisions considered. The improvement 
increases with the precision, which is as expected because our technique has 
complexity $n$ whereas computing arcsines has complexity worse than quadratic.

\begin{table}
\begin{tabular}{|c|c|c|c|}
  \hline
  $n$  & $\comp$ Toffolis & arcsine Toffolis & improvement factor \\
  \hline
  $17$ & $17$ & $4872$ & $286$ \\
  $23$ & $23$ & $7784$ & $338$ \\
  $30$ & $30$ & $11264$ & $375$ \\
  \hline 
\end{tabular}
\caption{Comparison between our $\comp$ routine and the cost of calculating an 
arcsine on a quantum computer, which is the main cost of executing $\rot$. From 
the left, the columns present a value for $n$, the number of Toffoli gates 
needed to execute $\comp$ (the complexity for our approach), the lowest number 
of Toffoli gates reported in Table~II of~\cite{HRS18} for target precision equal 
to approximately $2^{-n}$, and the factor by which our work improves on the 
complexity of the previous approach.}
\label{tab:compare_to_Grover}
\end{table}

The factor of $n^2$ in the complexity of \cite{HRS18} originates from the 
complexity for multiplication. In principle the asymptotic complexity for 
computing an arcsine could be made close to $\order{n}$ by using more advanced 
methods of multiplication~\cite{BZ11}. However, that complexity includes a very 
large constant factor, so those methods would only be useful for unrealistically 
high precision. Our method gives $n$ complexity with a constant factor of one, 
so is more efficient than any method for calculating arcsines to any precision.

Before concluding, we briefly discuss errors due to approximation. For the 
linear coefficients problem, the output of our algorithm is exactly proportional 
to $\sum_\ell \alpha_\ell^{(n)} \ket{\ell}$. Grover's approach can only 
approximate this state due to imprecision in the arcsine calculation and single 
qubit rotations. For the root coefficients problem, our algorithm incurs error 
because we can apply $\unif^{-1}$ only approximately. When comparing to the 
target state with the exact coefficients $\alpha_\ell$ or $\sqrt{\alpha_\ell}$ 
there is additional error because of the finite-precision approximation of 
$\alpha_\ell$. This error scales as $2^{-n}$.


In conclusion, we have devised a modification to Grover state preparation that 
avoids the need for a quantum computer to do arithmetic. Whereas Grover's 
original approach required the quantum computer to calculate a rotation angle as 
the arcsine of an input value, our approach eschews this step in favor of an 
inequality test between an input value and an even superposition of all possible 
values. The inequality test marks those parts of the superposition that are 
smaller than the input value, meaning that the number of marked items is 
proportional to the input value. Following amplitude amplification, the input 
value has thus been transduced to an amplitude. By replacing arithmetic with an 
inequality test, we make significant reductions to the complexity of Hamiltonian 
simulation as it would be performed in practice.

We expect our practical complexity reduction to have broad impact throughout 
quantum algorithms research. In particular, our algorithm can replace Grover's 
in the implementation of walk operators in quantum-walk-based algorithms and in 
the LCU technique. The improvement to LCU is likely to enable improvements to 
quantum algorithms for simulating quantum chemistry, which is a major potential 
application of quantum computing.

It may be possible to perform a proof-of-principle experiment of our state 
preparation algorithm using a noisy intermediate-scale quantum processor. By 
setting the precision of the oracle to be small (e.g.,~$n=2$ for an eight qubit 
demonstration) and skipping amplitude amplification, very few gates would be 
required. Grover state preparation using arcsines would be out of reach without 
large-scale error-corrected quantum computers.


\acknowledgments
We thank Richard Brent, M\'aria Kieferov\'a, Barry Sanders, and Beno\^it Valiron 
for helpful discussions. DWB thanks Craig Gidney for pointing out how to perform 
an inequality test based on the techniques in~\cite{CDKM04} and~\cite{Gid18}. 
DWB is funded by the Australian Government through the Australian Research 
Council (Grant No.\ DP160102426).

\bibliographystyle{apsrev4-1-fixed}
\bibliography{references}

\begin{thebibliography}{22}%
\makeatletter
\providecommand \@ifxundefined [1]{%
 \@ifx{#1\undefined}
}%
\providecommand \@ifnum [1]{%
 \ifnum #1\expandafter \@firstoftwo
 \else \expandafter \@secondoftwo
 \fi
}%
\providecommand \@ifx [1]{%
 \ifx #1\expandafter \@firstoftwo
 \else \expandafter \@secondoftwo
 \fi
}%
\providecommand \natexlab [1]{#1}%
\providecommand \enquote  [1]{``#1''}%
\providecommand \bibnamefont  [1]{#1}%
\providecommand \bibfnamefont [1]{#1}%
\providecommand \citenamefont [1]{#1}%
\providecommand \href@noop [0]{\@secondoftwo}%
\providecommand \href [0]{\begingroup \@sanitize@url \@href}%
\providecommand \@href[1]{\@@startlink{#1}\@@href}%
\providecommand \@@href[1]{\endgroup#1\@@endlink}%
\providecommand \@sanitize@url [0]{\catcode `\\12\catcode `\$12\catcode
  `\&12\catcode `\#12\catcode `\^12\catcode `\_12\catcode `\%12\relax}%
\providecommand \@@startlink[1]{}%
\providecommand \@@endlink[0]{}%
\providecommand \url  [0]{\begingroup\@sanitize@url \@url }%
\providecommand \@url [1]{\endgroup\@href {#1}{\urlprefix }}%
\providecommand \urlprefix  [0]{URL }%
\providecommand \Eprint [0]{\href }%
\providecommand \doibase [0]{http://dx.doi.org/}%
\providecommand \selectlanguage [0]{\@gobble}%
\providecommand \bibinfo  [0]{\@secondoftwo}%
\providecommand \bibfield  [0]{\@secondoftwo}%
\providecommand \translation [1]{[#1]}%
\providecommand \BibitemOpen [0]{}%
\providecommand \bibitemStop [0]{}%
\providecommand \bibitemNoStop [0]{.\EOS\space}%
\providecommand \EOS [0]{\spacefactor3000\relax}%
\providecommand \BibitemShut  [1]{\csname bibitem#1\endcsname}%
\let\auto@bib@innerbib\@empty
\bibitem [{\citenamefont {Grover}(2000)}]{Gro00}%
  \BibitemOpen
  \bibfield  {author} {\bibinfo {author} {\bibfnamefont {L.~K.}\ \bibnamefont
  {Grover}},\ }\href {\doibase 10.1103/PhysRevLett.85.1334} {\bibfield
  {journal} {\bibinfo  {journal} {Phys. Rev. Lett.}\ }\textbf {\bibinfo
  {volume} {85}},\ \bibinfo {pages} {1334} (\bibinfo {year}
  {2000})}\BibitemShut {NoStop}%
\bibitem [{\citenamefont {Grover}(1997)}]{Gro97}%
  \BibitemOpen
  \bibfield  {author} {\bibinfo {author} {\bibfnamefont {L.~K.}\ \bibnamefont
  {Grover}},\ }\href {\doibase 10.1103/PhysRevLett.79.325} {\bibfield
  {journal} {\bibinfo  {journal} {Phys. Rev. Lett.}\ }\textbf {\bibinfo
  {volume} {79}},\ \bibinfo {pages} {325} (\bibinfo {year} {1997})}\BibitemShut
  {NoStop}%
\bibitem [{\citenamefont {Childs}(2009)}]{Chi09}%
  \BibitemOpen
  \bibfield  {author} {\bibinfo {author} {\bibfnamefont {A.~M.}\ \bibnamefont
  {Childs}},\ }\href {\doibase 10.1007/s00220-009-0930-1} {\bibfield  {journal}
  {\bibinfo  {journal} {Communications in Mathematical Physics}\ }\textbf
  {\bibinfo {volume} {294}},\ \bibinfo {pages} {581} (\bibinfo {year}
  {2009})}\BibitemShut {NoStop}%
\bibitem [{\citenamefont {Berry}\ and\ \citenamefont {Childs}(2012)}]{BC12}%
  \BibitemOpen
  \bibfield  {author} {\bibinfo {author} {\bibfnamefont {D.~W.}\ \bibnamefont
  {Berry}}\ and\ \bibinfo {author} {\bibfnamefont {A.~M.}\ \bibnamefont
  {Childs}},\ }\href
  {http://www.rintonpress.com/xxqic12/qic-12-12/0029-0062.pdf} {\bibfield
  {journal} {\bibinfo  {journal} {Quantum Information and Computation}\
  }\textbf {\bibinfo {volume} {12}},\ \bibinfo {pages} {0029} (\bibinfo {year}
  {2012})}\BibitemShut {NoStop}%
\bibitem [{\citenamefont {Low}\ and\ \citenamefont {Chuang}(2017)}]{LC17}%
  \BibitemOpen
  \bibfield  {author} {\bibinfo {author} {\bibfnamefont {G.~H.}\ \bibnamefont
  {Low}}\ and\ \bibinfo {author} {\bibfnamefont {I.~L.}\ \bibnamefont
  {Chuang}},\ }\href {\doibase 10.1103/PhysRevLett.118.010501} {\bibfield
  {journal} {\bibinfo  {journal} {Phys. Rev. Lett.}\ }\textbf {\bibinfo
  {volume} {118}},\ \bibinfo {pages} {010501} (\bibinfo {year}
  {2017})}\BibitemShut {NoStop}%
\bibitem [{\citenamefont {Kothari}(2014)}]{Kot14}%
  \BibitemOpen
  \bibfield  {author} {\bibinfo {author} {\bibfnamefont {R.}~\bibnamefont
  {Kothari}},\ }\emph {\bibinfo {title} {Efficient Algorithms in Quantum Query
  Complexity}},\ \href {http://hdl.handle.net/10012/8625} {Ph.D. thesis},\
  \bibinfo  {school} {University of Waterloo}, \bibinfo {address} {Waterloo,
  Ontario} (\bibinfo {year} {2014})\BibitemShut {NoStop}%
\bibitem [{\citenamefont {Berry}\ \emph
  {et~al.}(2015{\natexlab{a}})\citenamefont {Berry}, \citenamefont {Childs},\
  and\ \citenamefont {Kothari}}]{BCK15}%
  \BibitemOpen
  \bibfield  {author} {\bibinfo {author} {\bibfnamefont {D.~W.}\ \bibnamefont
  {Berry}}, \bibinfo {author} {\bibfnamefont {A.~M.}\ \bibnamefont {Childs}},\
  and\ \bibinfo {author} {\bibfnamefont {R.}~\bibnamefont {Kothari}},\ }in\
  \href {\doibase 10.1109/FOCS.2015.54} {\emph {\bibinfo {booktitle} {2015 IEEE
  56th Annual Symposium on Foundations of Computer Science}}}\ (\bibinfo {year}
  {2015})\ pp.\ \bibinfo {pages} {792--809}\BibitemShut {NoStop}%
\bibitem [{\citenamefont {Berry}\ \emph
  {et~al.}(2015{\natexlab{b}})\citenamefont {Berry}, \citenamefont {Childs},
  \citenamefont {Cleve}, \citenamefont {Kothari},\ and\ \citenamefont
  {Somma}}]{BCC+15}%
  \BibitemOpen
  \bibfield  {author} {\bibinfo {author} {\bibfnamefont {D.~W.}\ \bibnamefont
  {Berry}}, \bibinfo {author} {\bibfnamefont {A.~M.}\ \bibnamefont {Childs}},
  \bibinfo {author} {\bibfnamefont {R.}~\bibnamefont {Cleve}}, \bibinfo
  {author} {\bibfnamefont {R.}~\bibnamefont {Kothari}},\ and\ \bibinfo {author}
  {\bibfnamefont {R.~D.}\ \bibnamefont {Somma}},\ }\href {\doibase
  10.1103/PhysRevLett.114.090502} {\bibfield  {journal} {\bibinfo  {journal}
  {Phys. Rev. Lett.}\ }\textbf {\bibinfo {volume} {114}},\ \bibinfo {pages}
  {090502} (\bibinfo {year} {2015}{\natexlab{b}})}\BibitemShut {NoStop}%
\bibitem [{\citenamefont {Childs}\ \emph {et~al.}(2017)\citenamefont {Childs},
  \citenamefont {Kothari},\ and\ \citenamefont {Somma}}]{CKS17}%
  \BibitemOpen
  \bibfield  {author} {\bibinfo {author} {\bibfnamefont {A.~M.}\ \bibnamefont
  {Childs}}, \bibinfo {author} {\bibfnamefont {R.}~\bibnamefont {Kothari}},\
  and\ \bibinfo {author} {\bibfnamefont {R.~D.}\ \bibnamefont {Somma}},\ }\href
  {\doibase 10.1137/16m1087072} {\bibfield  {journal} {\bibinfo  {journal}
  {{SIAM} Journal on Computing}\ }\textbf {\bibinfo {volume} {46}},\ \bibinfo
  {pages} {1920} (\bibinfo {year} {2017})}\BibitemShut {NoStop}%
\bibitem [{\citenamefont {Low}\ and\ \citenamefont {Chuang}(2016)}]{LC16}%
  \BibitemOpen
  \bibfield  {author} {\bibinfo {author} {\bibfnamefont {G.~H.}\ \bibnamefont
  {Low}}\ and\ \bibinfo {author} {\bibfnamefont {I.~L.}\ \bibnamefont
  {Chuang}},\ }\href {http://arxiv.org/abs/1610.06546} {\bibfield  {journal}
  {\bibinfo  {journal} {arXiv:1610.06546}\ } (\bibinfo {year}
  {2016})}\BibitemShut {NoStop}%
\bibitem [{\citenamefont {Harrow}\ \emph {et~al.}(2009)\citenamefont {Harrow},
  \citenamefont {Hassidim},\ and\ \citenamefont {Lloyd}}]{HHL09}%
  \BibitemOpen
  \bibfield  {author} {\bibinfo {author} {\bibfnamefont {A.~W.}\ \bibnamefont
  {Harrow}}, \bibinfo {author} {\bibfnamefont {A.}~\bibnamefont {Hassidim}},\
  and\ \bibinfo {author} {\bibfnamefont {S.}~\bibnamefont {Lloyd}},\ }\href
  {\doibase 10.1103/PhysRevLett.103.150502} {\bibfield  {journal} {\bibinfo
  {journal} {Phys. Rev. Lett.}\ }\textbf {\bibinfo {volume} {103}},\ \bibinfo
  {pages} {150502} (\bibinfo {year} {2009})}\BibitemShut {NoStop}%
\bibitem [{\citenamefont {Wiebe}\ \emph {et~al.}(2012)\citenamefont {Wiebe},
  \citenamefont {Braun},\ and\ \citenamefont {Lloyd}}]{WBL12}%
  \BibitemOpen
  \bibfield  {author} {\bibinfo {author} {\bibfnamefont {N.}~\bibnamefont
  {Wiebe}}, \bibinfo {author} {\bibfnamefont {D.}~\bibnamefont {Braun}},\ and\
  \bibinfo {author} {\bibfnamefont {S.}~\bibnamefont {Lloyd}},\ }\href
  {\doibase 10.1103/PhysRevLett.109.050505} {\bibfield  {journal} {\bibinfo
  {journal} {Phys. Rev. Lett.}\ }\textbf {\bibinfo {volume} {109}},\ \bibinfo
  {pages} {050505} (\bibinfo {year} {2012})}\BibitemShut {NoStop}%
\bibitem [{\citenamefont {Clader}\ \emph {et~al.}(2013)\citenamefont {Clader},
  \citenamefont {Jacobs},\ and\ \citenamefont {Sprouse}}]{CJS13}%
  \BibitemOpen
  \bibfield  {author} {\bibinfo {author} {\bibfnamefont {B.~D.}\ \bibnamefont
  {Clader}}, \bibinfo {author} {\bibfnamefont {B.~C.}\ \bibnamefont {Jacobs}},\
  and\ \bibinfo {author} {\bibfnamefont {C.~R.}\ \bibnamefont {Sprouse}},\
  }\href {\doibase 10.1103/PhysRevLett.110.250504} {\bibfield  {journal}
  {\bibinfo  {journal} {Phys. Rev. Lett.}\ }\textbf {\bibinfo {volume} {110}},\
  \bibinfo {pages} {250504} (\bibinfo {year} {2013})}\BibitemShut {NoStop}%
\bibitem [{\citenamefont {Scherer}\ \emph {et~al.}(2017)\citenamefont
  {Scherer}, \citenamefont {Valiron}, \citenamefont {Mau}, \citenamefont
  {Alexander}, \citenamefont {van~den Berg},\ and\ \citenamefont
  {Chapuran}}]{SVM+17}%
  \BibitemOpen
  \bibfield  {author} {\bibinfo {author} {\bibfnamefont {A.}~\bibnamefont
  {Scherer}}, \bibinfo {author} {\bibfnamefont {B.}~\bibnamefont {Valiron}},
  \bibinfo {author} {\bibfnamefont {S.-C.}\ \bibnamefont {Mau}}, \bibinfo
  {author} {\bibfnamefont {S.}~\bibnamefont {Alexander}}, \bibinfo {author}
  {\bibfnamefont {E.}~\bibnamefont {van~den Berg}},\ and\ \bibinfo {author}
  {\bibfnamefont {T.~E.}\ \bibnamefont {Chapuran}},\ }\href
  {http://dx.doi.org/10.1007/s11128-016-1495-5} {\bibfield  {journal} {\bibinfo
   {journal} {Quantum Information Processing}\ }\textbf {\bibinfo {volume}
  {16}} (\bibinfo {year} {2017})}\BibitemShut {NoStop}%
\bibitem [{\citenamefont {H{\"a}ner}\ \emph {et~al.}(2018)\citenamefont
  {H{\"a}ner}, \citenamefont {Roetteler},\ and\ \citenamefont {Svore}}]{HRS18}%
  \BibitemOpen
  \bibfield  {author} {\bibinfo {author} {\bibfnamefont {T.}~\bibnamefont
  {H{\"a}ner}}, \bibinfo {author} {\bibfnamefont {M.}~\bibnamefont
  {Roetteler}},\ and\ \bibinfo {author} {\bibfnamefont {K.~M.}\ \bibnamefont
  {Svore}},\ }\href {https://arxiv.org/abs/1805.12445} {\bibfield  {journal}
  {\bibinfo  {journal} {arXiv:1805.12445}\ } (\bibinfo {year}
  {2018})}\BibitemShut {NoStop}%
\bibitem [{\citenamefont {Brassard}\ \emph {et~al.}(2002)\citenamefont
  {Brassard}, \citenamefont {H\o{}yer}, \citenamefont {Mosca},\ and\
  \citenamefont {Tapp}}]{BHMT02}%
  \BibitemOpen
  \bibfield  {author} {\bibinfo {author} {\bibfnamefont {G.}~\bibnamefont
  {Brassard}}, \bibinfo {author} {\bibfnamefont {P.}~\bibnamefont {H\o{}yer}},
  \bibinfo {author} {\bibfnamefont {M.}~\bibnamefont {Mosca}},\ and\ \bibinfo
  {author} {\bibfnamefont {A.}~\bibnamefont {Tapp}},\ }in\ \href {\doibase
  10.1090/conm/305} {\emph {\bibinfo {booktitle} {Quantum Computation and
  Information}}},\ \bibinfo {series} {Contemporary Mathematics}, Vol.\ \bibinfo
  {volume} {305},\ \bibinfo {editor} {edited by\ \bibinfo {editor}
  {\bibfnamefont {S.~J.}\ \bibnamefont {Lomonaco}}\ and\ \bibinfo {editor}
  {\bibfnamefont {H.~E.}\ \bibnamefont {Brandt}}}\ (\bibinfo  {publisher}
  {American Mathematical Society},\ \bibinfo {year} {2002})\BibitemShut
  {NoStop}%
\bibitem [{\citenamefont {Babbush}\ \emph {et~al.}(2018)\citenamefont
  {Babbush}, \citenamefont {Gidney}, \citenamefont {Berry}, \citenamefont
  {Wiebe}, \citenamefont {McClean}, \citenamefont {Paler}, \citenamefont
  {Fowler},\ and\ \citenamefont {Neven}}]{BGB+18}%
  \BibitemOpen
  \bibfield  {author} {\bibinfo {author} {\bibfnamefont {R.}~\bibnamefont
  {Babbush}}, \bibinfo {author} {\bibfnamefont {C.}~\bibnamefont {Gidney}},
  \bibinfo {author} {\bibfnamefont {D.~W.}\ \bibnamefont {Berry}}, \bibinfo
  {author} {\bibfnamefont {N.}~\bibnamefont {Wiebe}}, \bibinfo {author}
  {\bibfnamefont {J.}~\bibnamefont {McClean}}, \bibinfo {author} {\bibfnamefont
  {A.}~\bibnamefont {Paler}}, \bibinfo {author} {\bibfnamefont
  {A.}~\bibnamefont {Fowler}},\ and\ \bibinfo {author} {\bibfnamefont
  {H.}~\bibnamefont {Neven}},\ }\href {https://arxiv.org/abs/1805.03662}
  {\bibfield  {journal} {\bibinfo  {journal} {arXiv:1805.03662}\ } (\bibinfo
  {year} {2018})}\BibitemShut {NoStop}%
\bibitem [{\citenamefont {Gily{\'e}n}\ \emph {et~al.}(2018)\citenamefont
  {Gily{\'e}n}, \citenamefont {Su}, \citenamefont {Low},\ and\ \citenamefont
  {Wiebe}}]{GSLW18}%
  \BibitemOpen
  \bibfield  {author} {\bibinfo {author} {\bibfnamefont {A.}~\bibnamefont
  {Gily{\'e}n}}, \bibinfo {author} {\bibfnamefont {Y.}~\bibnamefont {Su}},
  \bibinfo {author} {\bibfnamefont {G.~H.}\ \bibnamefont {Low}},\ and\ \bibinfo
  {author} {\bibfnamefont {N.}~\bibnamefont {Wiebe}},\ }\href
  {http://arxiv.org/abs/1806.01838} {\bibfield  {journal} {\bibinfo  {journal}
  {arXiv:1806.01838}\ } (\bibinfo {year} {2018})}\BibitemShut {NoStop}%
\bibitem [{\citenamefont {Fowler}\ \emph {et~al.}(2012)\citenamefont {Fowler},
  \citenamefont {Mariantoni}, \citenamefont {Martinis},\ and\ \citenamefont
  {Cleland}}]{FMMC12}%
  \BibitemOpen
  \bibfield  {author} {\bibinfo {author} {\bibfnamefont {A.~G.}\ \bibnamefont
  {Fowler}}, \bibinfo {author} {\bibfnamefont {M.}~\bibnamefont {Mariantoni}},
  \bibinfo {author} {\bibfnamefont {J.~M.}\ \bibnamefont {Martinis}},\ and\
  \bibinfo {author} {\bibfnamefont {A.~N.}\ \bibnamefont {Cleland}},\ }\href
  {http://arxiv.org/abs/1208.0928} {\bibfield  {journal} {\bibinfo  {journal}
  {arXiv:1208.0928}\ } (\bibinfo {year} {2012})}\BibitemShut {NoStop}%
\bibitem [{\citenamefont {Gidney}(2018)}]{Gid18}%
  \BibitemOpen
  \bibfield  {author} {\bibinfo {author} {\bibfnamefont {C.}~\bibnamefont
  {Gidney}},\ }\href {\doibase 10.22331/q-2018-06-18-74} {\bibfield  {journal}
  {\bibinfo  {journal} {{Quantum}}\ }\textbf {\bibinfo {volume} {2}},\ \bibinfo
  {pages} {74} (\bibinfo {year} {2018})}\BibitemShut {NoStop}%
\bibitem [{\citenamefont {Cuccaro}\ \emph {et~al.}(2004)\citenamefont
  {Cuccaro}, \citenamefont {Draper}, \citenamefont {Kutin},\ and\ \citenamefont
  {Moulton}}]{CDKM04}%
  \BibitemOpen
  \bibfield  {author} {\bibinfo {author} {\bibfnamefont {S.~A.}\ \bibnamefont
  {Cuccaro}}, \bibinfo {author} {\bibfnamefont {T.~G.}\ \bibnamefont {Draper}},
  \bibinfo {author} {\bibfnamefont {S.~A.}\ \bibnamefont {Kutin}},\ and\
  \bibinfo {author} {\bibfnamefont {D.~P.}\ \bibnamefont {Moulton}},\ }\href
  {https://arxiv.org/abs/quant-ph/0410184} {\bibfield  {journal} {\bibinfo
  {journal} {arXiv:quant-ph/0410184}\ } (\bibinfo {year} {2004})}\BibitemShut
  {NoStop}%
\bibitem [{\citenamefont {Brent}\ and\ \citenamefont
  {Zimmermann}(2011)}]{BZ11}%
  \BibitemOpen
  \bibfield  {author} {\bibinfo {author} {\bibfnamefont {R.~P.}\ \bibnamefont
  {Brent}}\ and\ \bibinfo {author} {\bibfnamefont {P.}~\bibnamefont
  {Zimmermann}},\ }\href {https://arxiv.org/abs/1004.4710} {\emph {\bibinfo
  {title} {Modern computer arithmetic}}}\ (\bibinfo  {publisher} {Cambridge
  University Press},\ \bibinfo {address} {Cambridge, New York},\ \bibinfo
  {year} {2011})\BibitemShut {NoStop}%
\end{thebibliography}%
\end{document}